\documentclass{aa}  
\usepackage{amsmath,inputenx,nicefrac,mathrsfs}
\usepackage{xcolor}
\usepackage{graphicx,subcaption}
\usepackage{multicol,multirow,blindtext}
\usepackage{wasysym,longtable}
\usepackage{pstricks}
\usepackage{txfonts,url,hyperref}
\usepackage{nicefrac,cancel,dashbox}
\usepackage{tikz,tkz-euclide,longtable,marvosym}

\begin{document} 

\title{Neptune's ring arcs from VLT/SPHERE-IRDIS near-infrared observations\thanks{This study is based on observations made with ESO Telescopes at the La Silla
Paranal Observatory, Chile, under programme SPHERE ID~-\,097.C-~0336 (A).} }

\titlerunning{Neptune's ring arcs}
\authorrunning{D. Souami et al.}

\author{%
D. Souami \inst{1, 2},
S. Renner \inst{3, 4}, 
B. Sicardy \inst{1}, 
M. Langlois \inst{5}, 
B. Carry \inst{6},
P. Delorme \inst{7},
\and
P. Golaszewska \inst{8}
}%

\institute{LESIA UMR-8109, Observatoire de Paris, Universit\'e PSL, CNRS, Sorbonne Universit\'e, 
Univ. Paris Diderot, Sorbonne Paris Cit\'e, 
5 place Jules Janssen, 92195 Meudon, France
\and
naXys, University of Namur, 8 Rempart de la Vierge, Namur, B-5000, Belgium
   \and
   Universit\'e de Lille, Observatoire de Lille, 1 impasse de l’Observatoire, 59000 Lille, France  
     \and
IMCCE, Observatoire de Paris, CNRS UMR 8028, 77 Avenue Denfert Rochereau, 75014 Paris, France
         \and     
     Univ. Lyon, Univ. Lyon\,1, ENS de Lyon, CNRS, Centre de Recherche Astrophysique de Lyon UMR5574, F-69230, Saint-Genis-Laval, France
     \and
      Universit\'e de la C\^ote d'Azur, Observatoire de la C\^ote d'Azur, CNRS, Laboratoire Lagrange, France
      \and   
Univ. Grenoble Alpes, CNRS, IPAG, F-38000 Grenoble, France
         \and
 University of Toronto Scarborough, Toronto, ON M1C 1A4, Canada.
\\
             \email{damya.souami@obspm.fr ; souami@astro.utoronto.ca}
}

\date{Received 20 June 2021; accepted 15 October 2021}

 
\abstract
{%
Neptune's incomplete ring arcs have been stable since their discovery in 1984 by stellar occultation. Although these structures should be destroyed within a few months through differential Keplerian motion, imaging data over the past couple of decades has shown that these structures are persistent. 
}%
{%
We present here the first SPHERE near-infrared observations of Neptune's ring arcs taken at  2.2\,$\mu$m (BB-Ks) with the IRDIS camera at the Very Large Telescope in August 2016.
}%
{%
The images were aligned using the ephemerides of the satellite Proteus and were suitably co-added to enhance ring and satellite signals. 
}%
{%
We analyse high-angular resolution near-infrared images of Neptune’s ring arcs obtained in 2016 at the ESO VLT-UT3 with the adaptive-optics fed camera SPHERE-IRDIS. We derive here accurate mean motion values for the arcs and the nearby satellite Galatea. The trailing arcs Fraternit\'e and Egalit\'e are stable since they were last observed in 2007. Furthermore, we confirm the fading away of the leading arcs Courage and Libert\'e. Finally, we confirm the mismatch between the arcs' position and 42:43 inclined and eccentric corotation resonances with Galatea; thus demonstrating that no 42:43 corotation model works to explain the azimuthal confinement of the arcs' materiel.
}%
{%
}%
      \keywords{planets and satellites: individual: Neptune -- planets and satellites: rings -- celestial mechanics -- techniques: photometric.}
   \maketitle  
%

\section{Introduction\label{sec:introduction}}
A stellar occultation campaign on July 22\textsuperscript{nd}, 1984 yielded the first conclusive evidence of an incomplete ring-like structure around Neptune, inside the classical Roche limit (assuming a density of 1 for the ring material)  \citep{Roques1984,Sicardy85a,Hubbard86,Covault86}. A summary of the discovery of the arcs is given in \citep{Nicholson90}; the authors compiled the results of five occultations (between April~18\textsuperscript{th}, 1984 and August~20\textsuperscript{th}, 1985), as well as the 1989 {\it Voyager} 2 data confirming three distinct features.  
 
  Voyager 2 data revealed that the arcs are longitudinally confined over a 40$^\circ$ azimuthal range; they are embedded in the much fainter continuous Adams ring around Neptune \citep{Smith89} which is the outermost ring of the Neptunian system. 
  
  These incomplete rings have shown to be stable since their discovery in 1984 \citep{Hubbard86}, while they would be expected to scatter in a few months through differential Keplerian motion. The azimuthal confinement of the arc system \citep{Goldreich86,Porco91} was thought to be a consequence of the arcs being within a 42:43 corotation inclination resonance (hereafter CIR) forced by Galatea. However, observations in 1998 \citep{Dumas99,Sicardy99} showed a slight mismatch between the observed arc mean motion and its expected value in the framework of the 42:43 CIR model.
  
  Furthermore, adaptive-optics data obtained with the Keck telescope in 2002 and 2003 \citep{Depater05} as well as VLT-UT3 NACO data of 2007 \citep{Renner14} showed persisting two trailing arcs. They also showed that the brightness and longitudes of the arcs (Fraternit\'e, Egalit\'e, Libert\'e, and Courage) have significantly changed since Voyager \citep{Depater05,Renner14,DePater2018}.
  
Several theoretical models addressed the question of the arcs' stability and the aforementioned mean motion mismatch. For example, \cite{Namouni02} have shown that the 42:43 corotation eccentricity resonance (CER) could match the current arcs' semi-major axis and stabilise the system, if the arcs contain a small fraction of Galatea's mass~($m_G$). Assuming a mass $m_G = 2.1\times10^{21}$g for Galatea, the model which is very sensitive to the uncertainty in Galatea's eccentricity ($e_G$) gave a ring mass of $\sim0.23\,m_G$ ($0.002\,m_G$, {\it resp.}) for $e_G=10^{-4}$ ($e_G=10^{-6}$, {\it resp.}). 

Hereafter, we adopt the 42:43 nomenclature for simplicity reasons. While the CER is a first order resonance (a true 42:43), the CIR is a second order resonance (84:86). \footnote{Their respective corotation critical arguments write as follows $\Psi_{CER} = 43 \lambda - 42 \lambda_G - \varpi_G$ and $\Psi_{CIR}=~{2\,[ 43 \lambda - 42 \lambda_G - \Omega_G]}$. Where $\lambda$ ($\lambda_G$) is the particle's (Galatea's) longitude, and $\varpi_G$ and $\Omega_G$ are the longitudes of the periapse and the node of the Galatea's orbit, respectively.}

Alternatively, multiple small co-orbital satellites in a stable stationary configuration that generalises the Lagrangian points \citep{Renner04,Renner14} are able to efficiently confine the dusty arc material, provided the right azimuthal spacings and masses of those co-orbital objects are chosen. In the framework of that model, \cite{Winter2020} investigated the arc orbital evolution under the effect of solar radiation.

\cite{Showalter2017} presented a possible three-body resonance mechanism capable of confining ring material within the observed corotation sites. This scenario constrains the orbital semi-major axis of the arcs to fall within $\sim$10 meters of a three-body mean motion resonance, which involves the two nearby inner satellites, Galatea and Larissa.

Finally, \cite{Showalter19} recently discovered a seventh inner moon of Neptune, Hippocamp, proving that the Neptunian system has not fully revealed itself. Furthermore, in the context of the possible Trident mission (preselected in NASA's Discovery Programme) targeting the Neptune-Triton system as well as a possible future ESA-NASA mission towards the Uranian and Neptunian systems; the next couple of decades could revolutionise our understanding of these systems.
   
   In this paper, we report on astro-photometric measurements of the arcs are obtained at the VLT- UT3 with the SPHERE-IRDIS instrument fed by its extreme adaptive optics system (SAXO) obtained on August 23\textsuperscript{rd}, 2016. The IRDIS classical imaging mode has also been used for several asteroid studies, leading to shape reconstruction \citep{Viikinkoski2015,Marsset2017}. These studies largely benefit from the high resolution and high Strehl providing much more detailed images than previous AO corrected images from other instruments. 

The paper is organised as follows. In section \ref{sec:Obs}, we present our observational data and the methods used for the analysis. Section~\ref{sec:Results} pertains to the photometric profiles of the arcs and the mean motion of the arcs and the satellite Galatea. Finally, we draw our final conclusions in Section \ref{sec:Conclusions}. 

\section{Observations\label{sec:Obs}}
 We used the high-angular resolution adaptive-optics Spectro-Polarimetric High-contrast Exoplanet REsearch (SPHERE) instrument installed on the Very Large Telescope VLT-UT3 at the European Southern Observatory (ESO), with the InfraRed Dual-band Imager and Spectrograph (IRDIS) \citep{Dohlen08,Beuzit2019}, to image the Neptunian system (ring arcs and moons) on August 23\textsuperscript{rd}, 2016. 
  
  IRDIS Classical imaging mode \citep{Langlois2010} provides simultaneous images recorded in the same broadband filter on two distinct detector areas. 
 The data were acquired using this classical mode of SPHERE, in the Ks broad-band filter centred at 2.2\,$\mu$m, which corresponds to a strong absorption in the methane spectrum, hence reducing the otherwise overwhelming scattered light from Neptune's atmosphere.
 
  In broad-band-Ks (BB-Ks), the instrument exhibits a fairly high thermal background from thermal emission reaching 60 to 200 photon s$^{-1}$ pixel$^{-1}$ \citep{Beuzit2019}, which renders nearly impossible the classical reduction procedures to identify faint objects such as small Neptunian satellites as well as the rings and arcs features. 
  
   IRDIS Dual-band Imager in classical imaging mode provides a FOV of $11"\times12.5"$ with a pixel size of $(12.255~\pm~0.009)$\,mas/pixel on the sky \citep{Maire2016}; this corresponds to $\sim 257$\,km/pixel at Neptune's geocentric distance of 28.959\,AU (cf. Table~\ref{tab:obsDataInfo}).

 We acquired field-stabilised data between 03:09:30.9281 and 07:58:36.2025 UT, on 2016-08-23. This acquisition mode uses a specific derotator rotation law to maintain the same field of view orientation during the observing sequence. 
 
\begin{figure}[htb]
   \centering
 \includegraphics[width=9.5cm]{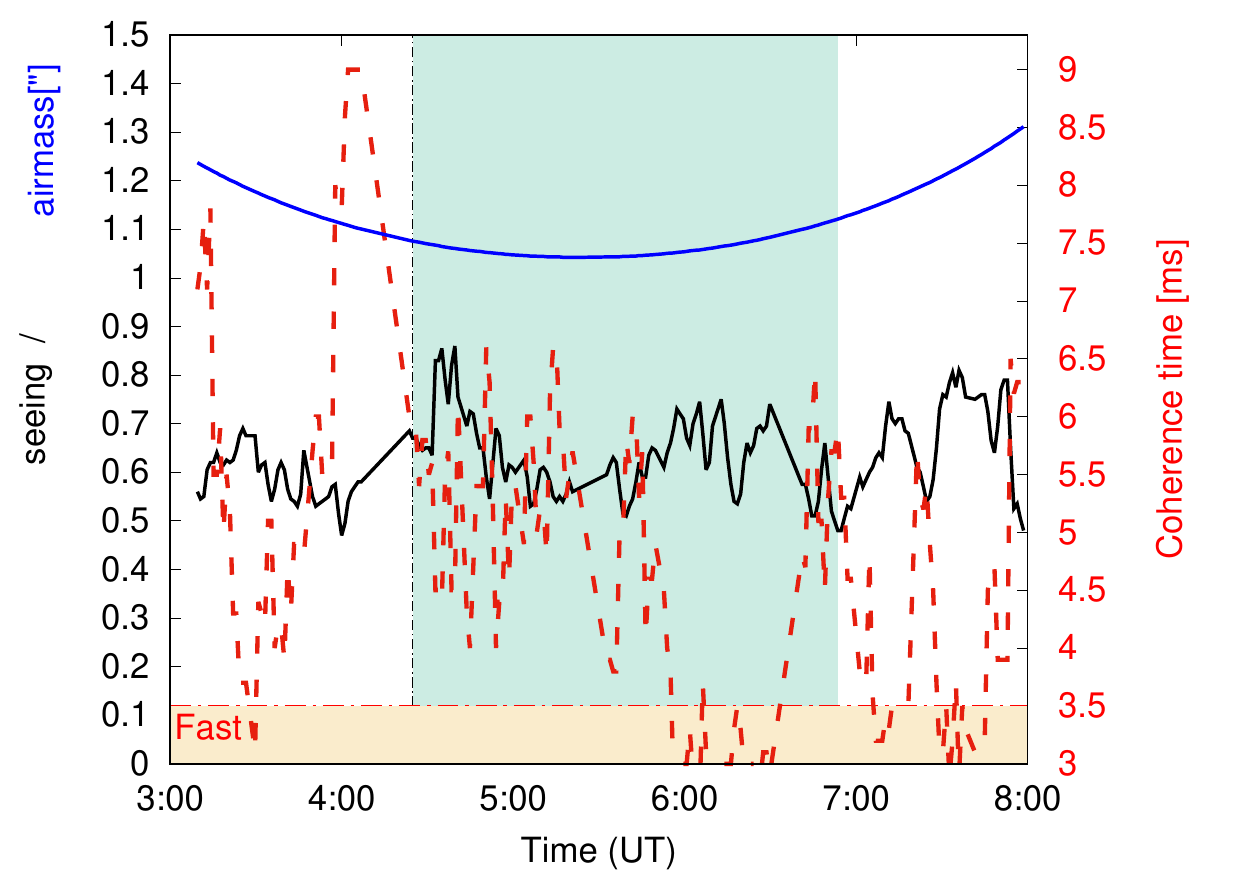}
       \caption{\label{fig:ObsConditions} The observing conditions (airmass, seeing, and coherence time) at Paranal during the observation run (between 03:09:30.9281 and 07:58:36.2025 UT on Aug. 23\textsuperscript{rd}, 2016). The large fluctuations in coherence time $\tau_0$ of atmospheric turbulence show that some images cannot be used, because of small $\tau_0$ values that is associated with fast atmospheric turbulence (the shaded light red horizontal band for which $\tau_0<3.5\,$ms). The green shaded area represents the time window from which the final images used in our analysis were extracted (see text for explanation). These images correspond to the lowest airmass observing window. 
}        
   \end{figure}

  Figure \ref{fig:ObsConditions} shows the details on the atmospheric conditions (seeing, airmass, and coherence time $\tau_0$) during the entire observational run, together with the time window of frames selected for the photometric analysis (green shaded area). Although the seeing remains fairly stable throughout the observation, about 27.4\% of our data (57 frames) were acquired at times associated with non-negligible atmospheric turbulence and thus not used for our photometric analysis. The threshold defined by ESO as associated with fast wind and/or fast turbulence conditions (\textit{i.e.} turbulence evolving on timescales smaller than 3.5 ms) is defined by low coherence times, \textit{i.e.} $\tau\le3.5\,$ms. These constraints combined with the instability of the AO tracking (see section \ref{sec:Analysis} for details), the highest quality data that we have selected and retained for the photometric analysis are obtained at the smallest airmass ($<1.1$).
      
\begin{table}[htbp]
\begin{center}
  \begin{tabular}{ l c } 
    \hline
       &  August 23\textsuperscript{rd}, 2016    \\
     \hline 
     \hline
Ref. frame UT time (hr:min:sec) & 05:33:51.9581 \\ 
MJD epoch of ref. frame  & 57623.23185137 \\ 
Geocentric distance (AU) & 28.959  \\
Heliocentric distance (AU) & 29.954  \\
Phase angle (deg) & 0.35138  \\
$B$ (deg) &-26.133 \\ 
$P$ (deg) &  326.237 \\ 
$U$ (deg) & 318.808 \\ 
\hline
Number of used images & 68  \\
Individual exposure time (sec) & 64  \\
\hline
Scale (mas/pixel)  &  $12.255 \pm 0.009$ \\
\hline                                              
  \end{tabular}
 \caption{\label{tab:obsDataInfo}Circumstances of observation: 
The epoch given is the time of the reference frame used to produce the equivalent width profile (cf. section \ref{subsec:Photometry}). 
 The geocentric and heliocentric distances, as well as the phase angle are retrieved from the Rings Node of NASA's Planetary Data System (\url{http://pds-rings.seti.org}). The angles $B$, $P$, and $U$ are the ring's opening angle to Earth, Neptune's pole orientation, and the longitude of Earth measured in the ring plane from the J2000.0 ascending node of Neptune's equatorial plane, respectively. The pixel scale value is that given by \cite{Maire2016}.}
  \end{center}
\end{table}    
%
\subsection{Data reduction} %

 We used the SPHERE pipeline package to correct for flat field, bad pixels and thermal background due to instrument and sky. Because the thermal background varies with the instrument temperature and with the sky background, we selected background calibration images acquired using the same integration time and the same filter (BB-Ks, of 300\,nm width) as the science frames, but acquired at a different time, so they match better the conditions of the observations. Some uncorrected background leaks persist in our final K images due to thermal background fluctuations. The astrometrisation (centring, correction for anamorphism, scale, and true north) of the final data cubes are corrected following the SPHERE Data Center procedures \citep{Delorme2017}.
 
We have also corrected the true-north offset of the images, by rotating the images by the angle $P_{cor}=(-1.57\,\pm\,0.08)$~deg \citep{Maire2016},  where $P_{cor}$ is the position angle of the celestial north direction with respect to the frame columns and the position of Neptune’s centre.

\subsection{Astro-photometric analysis of the data\label{sec:Analysis}}
 Using the IRDIS HAWAII-2-RG near-infrared detector, we initially acquired 208\,exposures of 64\,seconds exposure time each. However, because of the aforementioned thermal background, passing clouds, and varying winds, only the best 68 images of the right panel were deemed of high enough SNR and used for our analysis (cf. Table~\ref{tab:obsDataInfo}).

\begin{figure*}[htb]
   \centering
   \includegraphics[width=6.5cm]{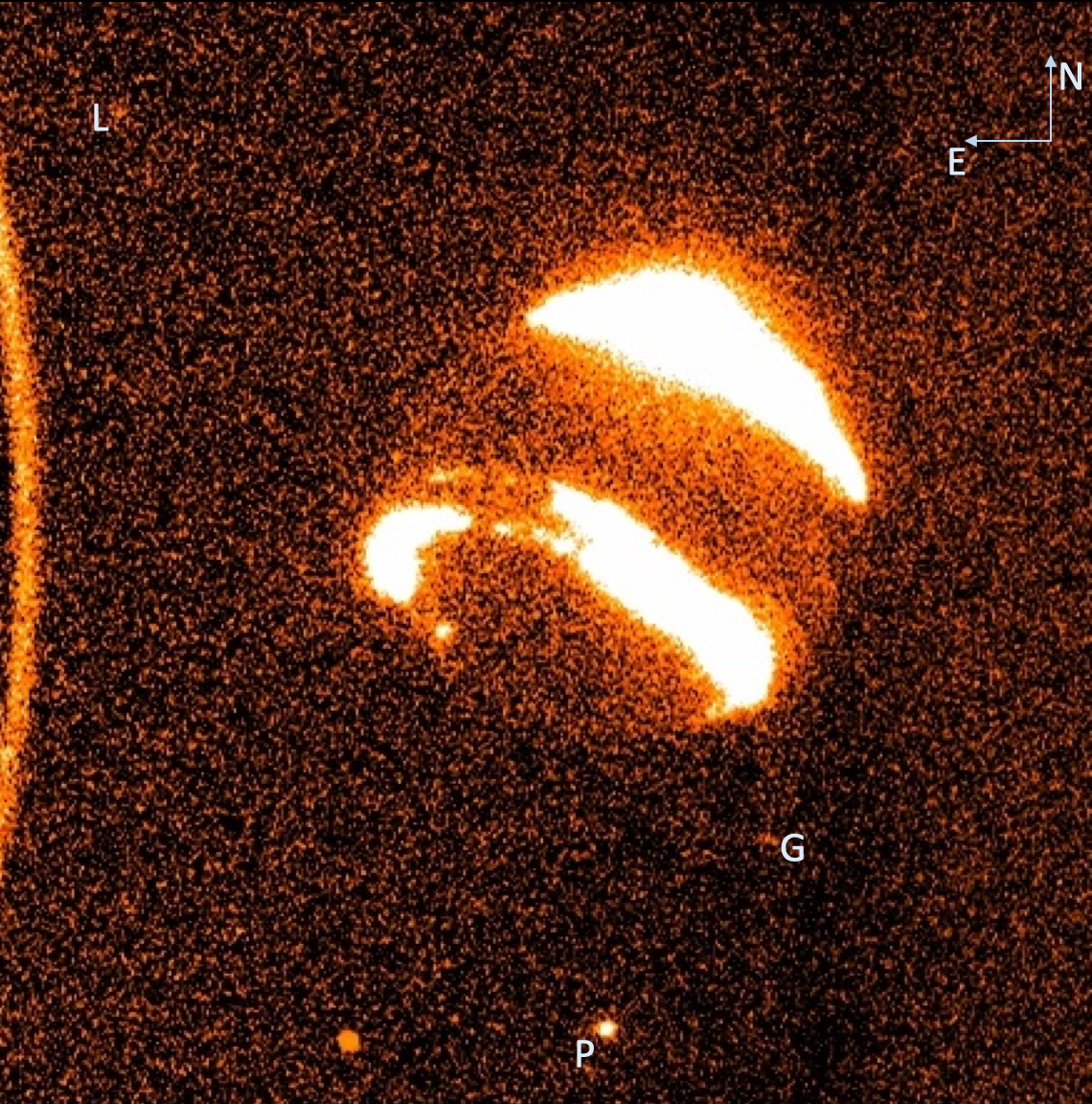}
  \hskip 2cm
 \includegraphics[width=7.5cm]{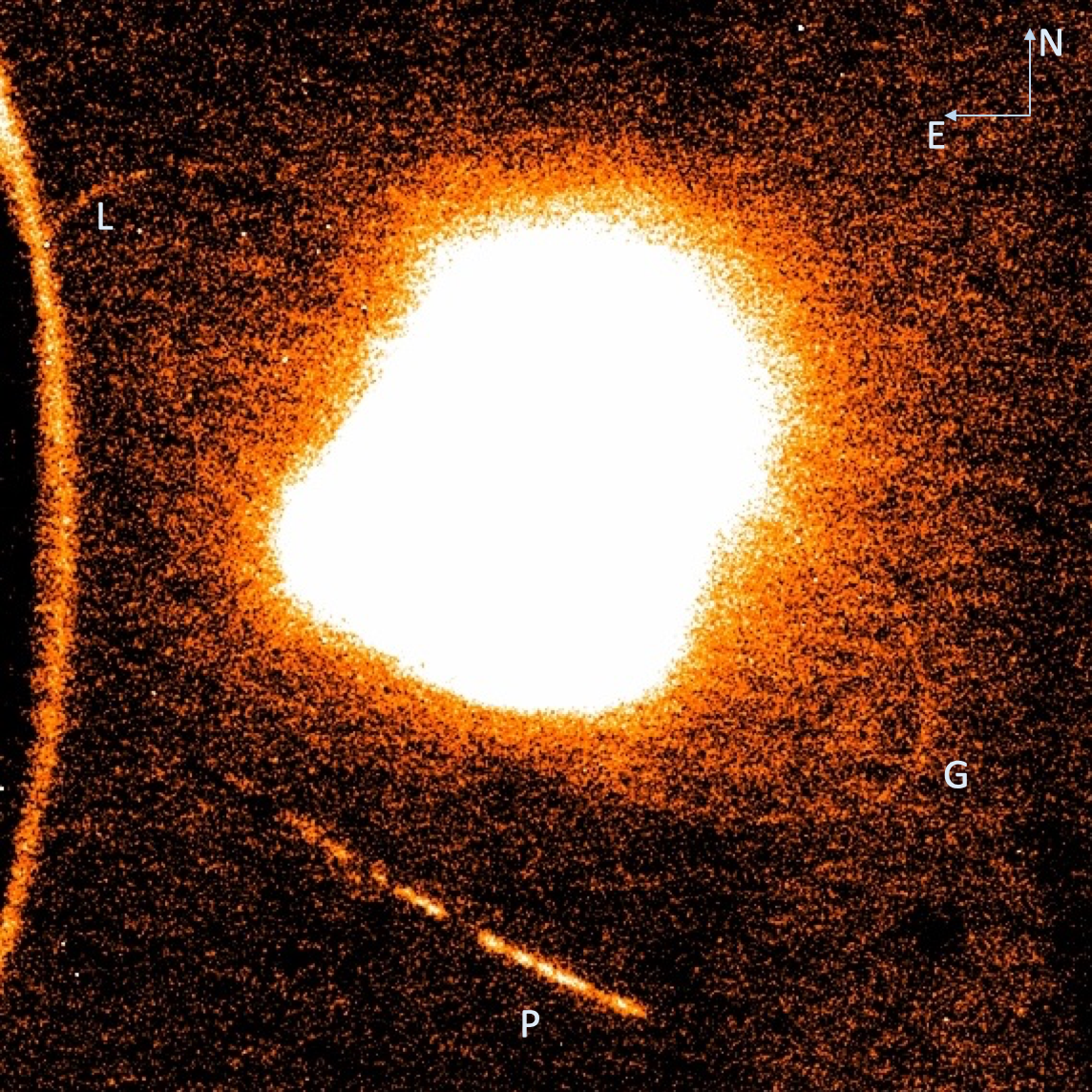}
       \caption{\label{fig:neptune_in_K}(left) Single 64 seconds exposure of Neptune in the BB-Ks band (2.182\,$\mu$m), 2016-08-23T05:12:52.2635\,UT August~23\textsuperscript{rd}, 2016, revealing the satellites Proteus (P), Galatea~(G), and Larissa (L). The frame is $5.39\times5.92$ arcsec$^2$ wide. 
 The point source is due to a bad pixel cluster which is difficult to properly interpolate.
 (right)~Image of Neptune's system obtained by co-adding 68 individual frames of the August~23\textsuperscript{rd}, 2016 data (72 minutes and 32 seconds total exposure time, between 04:24:50.5202 and 06:43:23.0627\,UT). The frame is $6.99\times6.99$ arcsec$^2$ wide. We can see Proteus (P) and Larissa (L) along their respective orbits during this time interval. We can also see the trail of Galatea (G) and the arcs, which evolve along the much fainter Adams ring. The bright arc on the left is due to thermal electronic noise.}        
\end{figure*}

In \cite{Renner14}, we were able to refine the determination of the pixel scale and the orientation on sky by comparing the positions of Triton and Proteus to their  expected relative positions. In this paper, given the fact that Proteus is the only satellite with sufficiently accurate astrometry and ephemerides, we had to operate under the following assumptions: {\it (i)} perfect ephemerides of the satellite Proteus which are used to retrieve Neptune's centre position, {\it (ii)}~the scale and orientation provided by the SPHERE-IRDIS consortium are correct. 
 The instability of the AO correction due to the observation of such extended object as Neptune (2.4" in diameter) used as the guide star, results in image motion that required a re-alignment. Such re-alignment is obtained by using Proteus' ephemerides \citep{Jacobson09} because the photo-centre of Neptune cannot be accurately determined (due to its non-uniformity and its extended nature). 
To centre the images, we first determined an approximate position (photo-centre) of Proteus on each image. From the ephemerides and the measured position of the photo-centre, we were able to retrieve a 'first order' position for Neptune's centre. We then iterated the analysis on the first order centred images to which we subtracted the median of the first centred cube files in which Neptune only is stationary, thus refining our determination of Proteus' photo-centre along its orbit, and therefore the position of Neptune's centre. This last step was repeated on this new cube of centred images, as this process allows the removal of the diffuse scattered light around Neptune, thus improving the determination of the satellite's photo-centre. Two iterations of the procedures are indispensable to achieve a sub-pixel accuracy for the satellite's and Neptune's positions. Examples of an individual frame as well as a final stack of the centred clean images are presented in Figure ~\ref{fig:neptune_in_K}.

The individual images were then projected onto Neptune's equatorial plane. The single projected frames were rotated taking into account the respective orbital motion and co-added to increase the signal from the ring arcs or a given satellite.
\section{Results\label{sec:Results}}
\subsection{Photometry\label{subsec:Photometry}}  
\begin{figure*}[htb]	   
    \begin{subfigure}[b]{0.5\textwidth}
            \begin{center}
\includegraphics[width=0.6\textwidth]{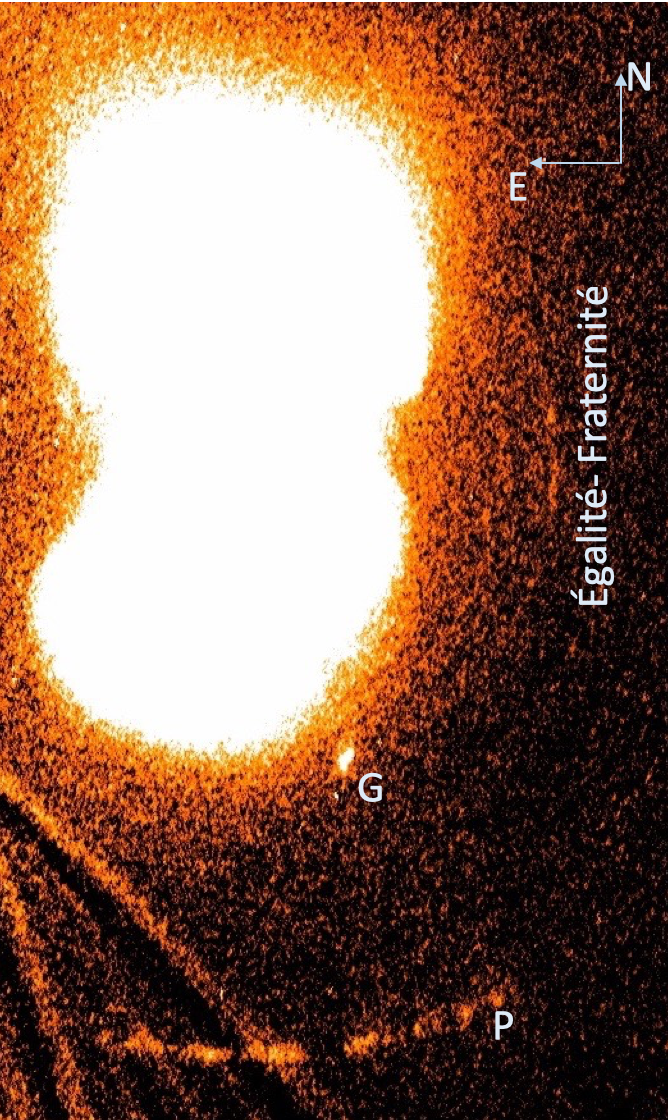}
      \caption{\label{subfig:projectedArcsSPHEREA}}
    \end{center}  
   \end{subfigure}%
    ~ 
    \begin{subfigure}[b]{0.5\textwidth}
            \begin{center}
   \includegraphics[width=1.0\textwidth]{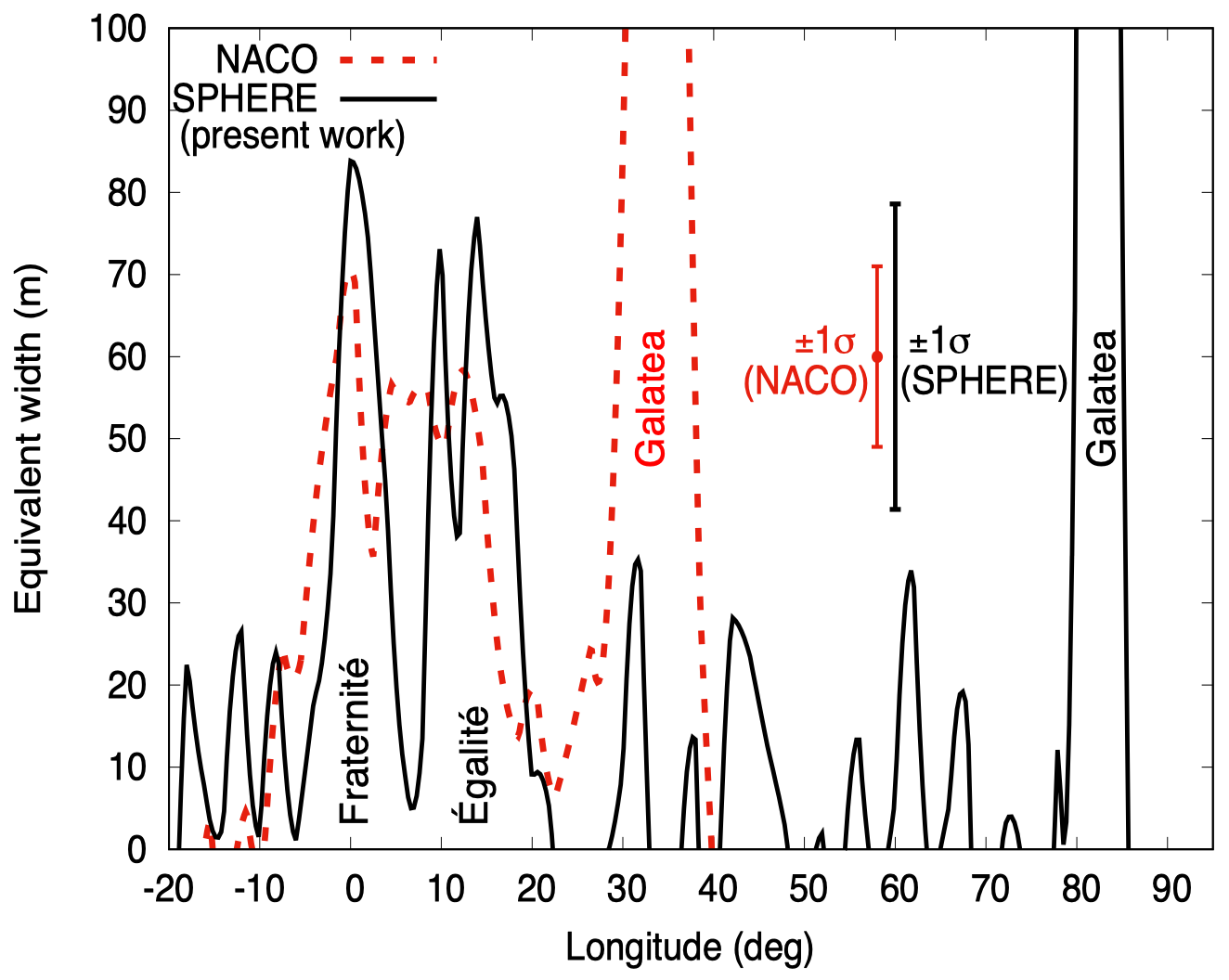}
         \caption{\label{fig:projectedArcsSPHEREB}}
       \end{center}  
    \end{subfigure}
      \caption{\label{fig:projectedArcsSPHERE} {\it (a)} The 68 projected and co-added images of Neptune's equatorial plane revealing material along the arcs Fraternit\'e and Egalit\'e, as well as the satellites Proteus (P) and Galatea~(G).  The frame is $7.1 \times 10.7\,$arcsec$^2$ wide. 
 {\it (b)} The equivalent width of the arcs (Fraternit\'e and Egalit\'e) vs. longitude
extracted from Figure~\ref{fig:projectedArcsSPHEREB}, at an angular resolution of 2$^\circ$. The X-axis origin is the longitude $L_{Fr}$ of the centre of the Fraternit\'e arc measured from J2000.0 ascending node, where $L_{Fr}=217.43\,$deg, at the reference epoch (cf.~Table \ref{tab:obsDataInfo}).} 
 \end{figure*}

 In what follows, we use only the best 68 selected projected frames, taken between 04:24:50.5202 and 06:43:23.0627 on August 23\textsuperscript{rd}, 2016 UT. These are co-added after correction for the arcs' mean motion, $820.11213$\,deg\,day$^{-1}$ \citep{Renner14}, to produce the panel on figure~\ref{subfig:projectedArcsSPHEREA} (total exposure time 72 minutes and 32 seconds).

Unlike the SPHERE data for which the field is depleted of stars, the 2007 NACO data \citep{Renner14} included a reference star which could be used for photometric calibration. We had to adapt to this additional constraint. 

 Using the same subset of SPHERE images with a good enough SNR for both Proteus and Galatea, we derotated these images at the respective mean motion values (320.765625$\pm$0.000001)\,deg\,day$^{-1}$ for Proteus and (839.661311$\pm$0.000005)\,deg\,day$^{-1}$ for Galatea \citep{Showalter19}. On these images, we measured the fluxes of the satellites using classical aperture photometry. We then compared the ratio of fluxes (Proteus/Galatea) for the NACO data \citep{Renner14} to that of the SPHERE data used in this paper, and found ratios of $0.12\pm0.01$ and $0.13\pm0.01$, respectively. The flux ratio of Proteus to Galatea is thus unchanged (at the $1\sigma$-error level).

Similarly to what was done in \citep{Renner14}, the brightness longitudinal profiles are given in equivalent width, \textit{i.e.} the width of a perfect Lambert diffuser that would reflect sunlight at the distance of Neptune. Following earlier studies of Neptune's rings (\cite{Smith89}; \cite{Porco95}),  the equivalent width is defined by 
 $E(\lambda)~=~\mu \int I(\lambda) / F(\lambda) dr,$ where $I(\lambda)$ denotes the observed flux reflected from the arcs, $\pi F(\lambda)$ the incident solar flux, $\lambda$~the wavelength at which observations were done, and $\mu$~denotes the cosine of the emission angle with respect to the ring-plane normal.
 
For our photometric analysis of the arcs' profile, we proceed as follows:
 \begin{enumerate}[\it i.]
\item We obtained the profiles of the arcs in ADU (Analog to Digital Units) by subtracting the sky background and Neptune's scattered light from the arcs' signal. To perform this operation, we integrated the ring region in the radial direction over 13 pixels. We selected circular annuli on both sides of the ring arcs, summed the pixels in the radial direction, and fitted a two-degree polynomial as a function of the longitude. The radial width of these inner and outer circular rings are $\sim$11 and 13 pixels, respectively. The average of the two polynomial fits was then subtracted from the arcs' flux. 
\item Using the same 68 individual projected frames as in Fig.~\ref{fig:neptune_in_K} which we rotate at Proteus' mean motion, we perform aperture photometry and extract the flux in ADU for the satellite.
\item We then normalise the rings' flux profile obtained in \textit{(i.)} by Proteus' flux obtained in \textit{(ii.)}.
\item Steps {\it(i.)} to {\it(iii.)} are then repeated on the NACO 2007 data taken at 2.2 $\mu$m (Ks band) \citep{Renner14}.
\item  We use the normalised (NACO 2007) ADU flux profile of the rings obtained in {\it(iv.)} and the equivalent width profile for these same rings \citep{Renner14} to derive the conversion factor between the normalised flux and the profile in equivalent width. We then apply this factor to the SPHERE 2016 data. This is permitted since both NACO and SPHERE data were acquired in Ks-BB at 2.2\,$\mu$m, and under the assumption of a fixed magnitude for Proteus (\textit{cf.} above).
 This approach is justified by the absence of a reference (calibration) star in the SPHERE 2016 data.
 \end{enumerate}

The equivalent width profiles of the ring arcs are presented in Fig.~\ref{fig:projectedArcsSPHEREB}. We detected the arcs Fraternit\'e and Egalit\'e, with flux densities and longitudinal extensions similar to those of the previous VLT-NACO observations \citep{Renner14} and Keck observations \citep{Depater05}. %

The trailing arcs Fraternit\'e and Egalit\'e are clearly separated in our data (see Figure \ref{subfig:projectedArcsSPHEREA}), which was not the case in the NACO data \citep{Renner14}. Despite the high thermal noise, we were able to draw  conclusions on the stability of the trailing arcs Fraternit\'e and \'Egalit\'e, and extract equivalent width profiles. Finally, this new data confirms that the leading arcs Courage and Libert\'e have faded away, as previous studies were leaning in that direction \citep{Dumas99,Depater05}. Accounting for the geocentric distance given in Table~\ref{tab:obsDataInfo}, we derived the equivalent width of the arcs Fraternit\'e and \'Egalit\'e at the epoch of observation at the 1$\sigma$ uncertainty level, and find that $E_{Fr}=(84\pm18)\,$m and $E_{Eg}=(76\pm18)$\,m, respectively. 

 For comparison, \cite{Renner14} and \cite{Depater05} found $E_{Fr}=(71\pm10)\,$m and $\sim$65\,m, respectively. In \citep{Renner14}, we reported that \'Egalit\'e was about $\sim$20\% fainter than Fraternit\'e from our 2007 VLT-NACO data (keeping in mind the pollution by Galatea of the arcs). On the other hand, \cite{Depater05} report that \'Egalit\'e was $\sim$17\% brighter than Fraternit\'e ($\sim3\sigma$) in their 2002 Keck II data, while its intensity had decreased to $\sim7\%$ below that of Fraternit\'e ($\sim2\sigma$) in 2003.    

To summarise, given the uncertainties on the equivalent width measurements, the arcs Fraternit\'e and \'Egalit\'e appear to be stable when compared to previous studies \citep{Dumas99,Depater05,Renner14}.

\subsection{Mean motions}
We improved the average mean motion of the satellite Galatea: \textit{(i)} at epoch 2016-08-23T04:25:22.5202\,UT, we derive the satellite's longitude $L=~ 259^\circ.58~\pm~0^\circ.30$, measured in the ring plane from the J2000.0 ascending node of Neptune's equatorial plane, \textit{(ii)}~using the reference longitudes measured by Voyager  \citep{Owen91} at epoch JD~2447757.0 ({\it i.e.} 1989 August~18\textsuperscript{th}, 12\,h) at Neptune, we derive the average mean motion of Galatea, $n_{G}~=~(839.66134\,\pm\,0.00003)$~deg~day$^{-1}$. This value is in agreement with that given in \citep{Showalter19}, $(839.661311~\pm~0.000005)\,$deg day$^{-1}$. 

  Furthermore, from the stacked image (Fig. \ref{subfig:projectedArcsSPHEREA}) with respect to the reference frame (see Table \ref{tab:obsDataInfo}), we derived a longitude $L_{Fr}~=~(217.43\pm0.30)$ deg for the trailing arc Fraternit\'e at this same reference epoch. We~used the position of the middle of this arc measured from Voyager data (251.88~deg at epoch JD 2447757.0, {\it i.e.} 1989 August 18,~12\textsuperscript{h}~UT) as given in \citep{Porco95}, and derived the following average mean motion for the arcs (Fraternit\'e and \'Egalit\'e):  $n_{arcs}~=(820.11178~\pm~0.00003)$~deg~day$^{-1}$ which is consistent with, yet more accurate than previous measurements ("solution 2" of \cite{Nicholson95,Depater05}), $820.1118~\pm~0.0001\,$deg day$^{-1}$. 
 We used our new measurement to compare $n_{arcs}$ with the mean motion of the 42:43 CIR, initially thought to confine dust within the ring to form stable arcs \citep{Goldreich86,Porco91}. 

\begin{table*}[htb]
\begin{center}
  \begin{tabular}{ l c c c c} 
    \hline
Resonance type &  drift in mean motion ($\Delta n$)& mismatch ($\Delta\,a$)& $\Delta L$ $^*$& \\
 & (deg/day) & (meters) & (deg) & \\
     \hline 
     \hline 
     42:43 CIR with Galatea & $n_{CIR}-n_{arcs}\approx(5.95\pm0.03)\,.\,10^{-3}$ & $a_{arcs}-a_{CIR} \approx305 \pm 2$ & $\sim(58.8\pm0.3)$ & \\
     42:43 CER with Galatea & $n_{CER}-n_{arcs}\approx(39.16\pm0.03)\,.\,10^{-3}$ & $a_{arcs}-a_{CER}\approx2004\,\pm\,2$ & $\sim(26.4\pm0.3)$ & \\
\hline                                              
  \end{tabular}
 \caption{\label{tab:testing_resonances}Investigating the 42:43 CIR and the 42:43 CER corotation models with Galatea.}
 \tablefoot{$^*$ $\Delta\,L$ represents the difference for the arcs' longitude (assuming a trapping in the CIR or the CER) over $\sim$27.01364~years (between the 1989 Voyager and our VLT SPHERE-IRDIS 2016~data).
}
  \end{center}
\end{table*}    
  
  The 42:43 CIR with Galatea creates 86 equally spaced corotation sites around Neptune, with a mean motion given by $n_{CIR}~= (42 n_G + \dot{\Omega}_G)/43$, where $\dot{\Omega}_G$ is Galatea's nodal precession rate. Using $\dot{\Omega}_G =-0.713675$ \,deg\,day$^{-1}$ \citep{Showalter19}, we find $n_{CIR} = 820.11773 \pm 0.00003$ deg\,day$^{-1}$. 
  This value of $ n_{CIR}$ is similar to previous mean motion measurements \citep{Nicholson95,Sicardy99,Dumas99,Dumas02,Depater05,Renner14}, showing that the arcs are not at the location of the 42:43 CIR with Galatea.  This mean motion drift translates into a mismatch in semi-major axis  $\Delta a=(305\pm2)\,$m (cf. Table \ref{tab:testing_resonances}). This mismatch is consistent with, yet more accurate than those reported in \citep{Sicardy99, Renner14}; the half-width of the CIR being $(250\pm100)\,$m \citep{Renner14}. 
   
Furthermore, we investigate the 42:43 corotation eccentric resonance (CER) with Galatea. This  resonance is defined as follows $n_{CER}~=~(42 n_G + \dot{\omega}_G)/43$, where $\dot{\omega}_G$ is Galatea's apsidal precession rate. With $\dot{\omega}_G =0.714282$ \,deg\,day$^{-1}$ \citep{Showalter19}, we find $n_{CER} = 820.15094 \pm 0.00003$ deg\,day$^{-1}$. 

\begin{table}[h]
\begin{center}
\begin{tabular}{llll}
\hline	
\multicolumn{4}{c}{Neptune's properties} \\
\hline
$GM$ 	& & 6\,835\,099.5 km$^3$\,s$^{-2}$ &  \\
$R$ 	& & 25225\,km	& \\
$J_2$	& & $3,408.43\times10^{-6}$ &\\
$J_4$ 	& & $-33.40 \times 10^{-6}$ &	\\
\hline
\hline
\multicolumn{4}{c}{Galatea's orbital elements and mass} \\
\hline
$i$ ($^\circ$)	& & $0.0231\pm 0.0091$ & \\
$e$ 	& & $0.00022\pm 0.00008 $ & \\
$M_G$ & & $(2.12\pm 0.08)\,10^{18}\,$kg & \\
\hline
\end{tabular}  
\end{center} 
\caption{\label{tab:parameters} Parameters used to explore the validity of the 42:43 CER and CIR models: Neptune's physical properties from \citep{Jacobson09} as well as derived orbital inclination and eccentricity of Galatea, derived by \cite{Showalter19}. Galatea's mass is that derived in \citep{Porco91}.}
\end{table}   

For both the 42:43 CIR and CER, we give in table~\ref{tab:testing_resonances} the corresponding mismatch values in semi-major axis, as well as the associated difference for the arcs' longitude ($\Delta L$) over $\sim$27.01364 years (between the 1989 Voyager and VLT SPHERE-IRDIS 2016 data). Note that these values are large and far from what we observe.
     
\begin{figure}[htb]
   \centering
   \includegraphics[width=8.5cm]{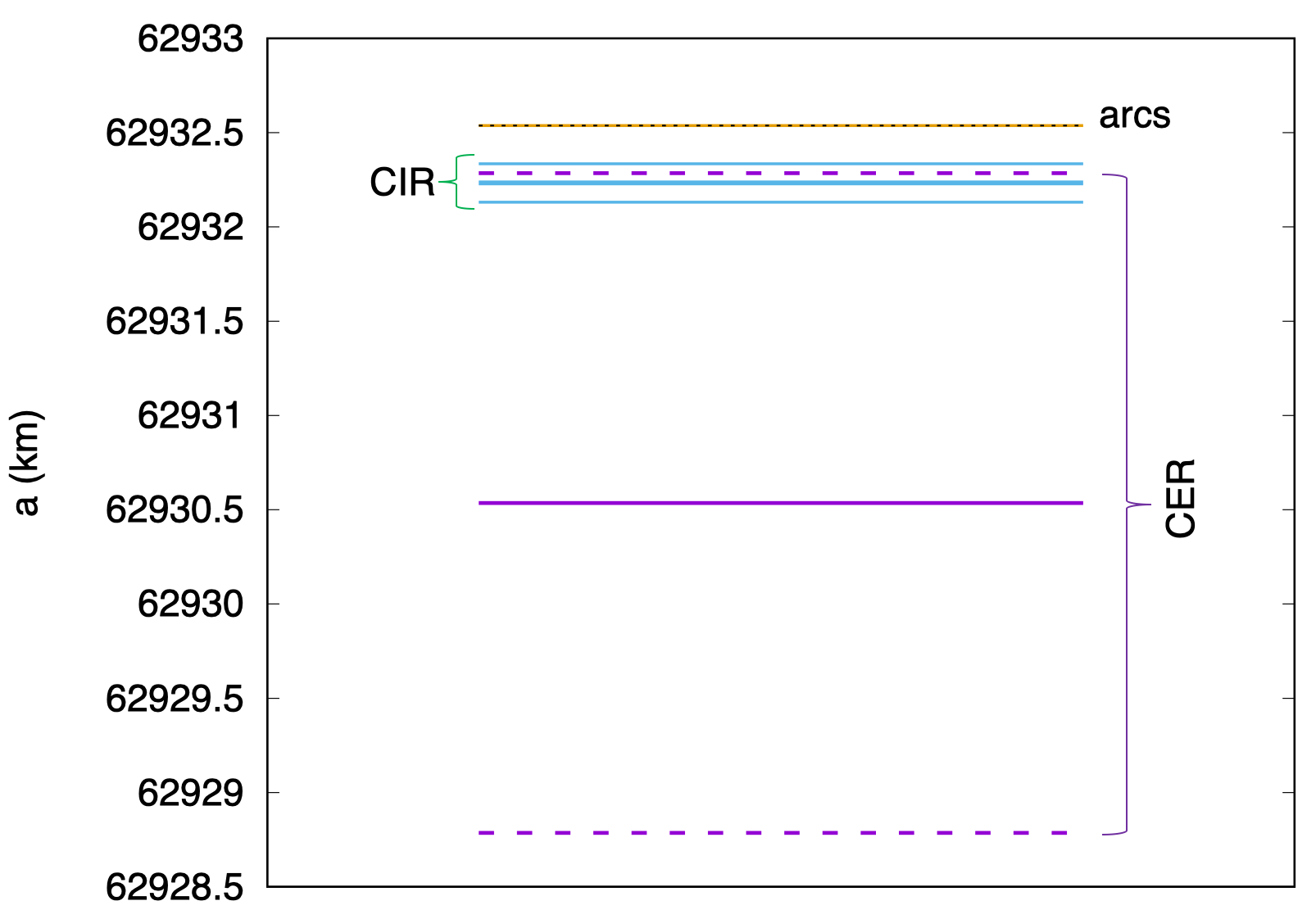}
       \caption{\label{fig:resonances} Graphical illustration of the arcs'  semi-major axis, as well as the radii and widths of the CIR and CER motion resonances 42:43 with Galatea. The orange continuous line depicts the geometrical orbital radius of the arcs, with uncertainties (in black dotted lines) smaller than the thickness of the orange line.
 The~CIR and CER locations and widths are represented in blue and purple, respectively. They represent the interval in semi-major axis where particles would be trapped by these resonances. This figure clearly shows the mismatch between the 42:43 CIR and the CER with Galatea; and the arc's position. The associated differences in longitude ($\Delta\,L$) in Table \ref{tab:testing_resonances} support this conclusion.
 }        
   \end{figure}

Figure \ref{fig:resonances} illustrates a graphic comparison of the semi-major axis of the arcs, as derived from the mean motion, as well the CER and CIR radii values derived in this work following the same approach as \cite{Foryta1996}. We used the parameters of Galatea and Neptune given in table \ref{tab:parameters}. In particular, this figure depicts clearly the widths ($\Delta\,a$) of each one of the CIR and CER resonances, {\it i.e.} the interval in semi-major axis where particles would be trapped by each one of these resonances. 

 From this analysis of the mismatch ($\Delta a$) and the resulting drifts in longitude $\Delta L$, we conclude that neither the CIR nor the CER 42:43 \citep{Namouni02} corotation models can explain the confinement of the arcs at their current locations. The valid model, so far, to explain the azimuthal spacing of the arcs is the model based on small co-orbital satellites \citep{Renner14}.   
 
 Finally, we test here the three-body mean motion resonance hypothesis  \citep{Showalter2017} of argument $(35\,n_G-~39\,n_{arcs}+~4\,n_L)$, which involves the satellites Galatea and Larissa. We use the mean motions for Galatea and the arcs derived in the work, and Larissa's mean motion $n_L=~(649.054085\pm 0.000004)$ deg\,day$^{-1}$ \citep{Showalter19}; we obtain a value $0.00382\,$deg\,day$^{-1}$. Further work is needed to confirm wether or not this three-body resonance is able to confine dust particles into corotation sites, with spacings compatible with the observations. 

\section{Conclusions\label{sec:Conclusions}}
We have analysed high-angular resolution near-infrared images of Neptune's ring arcs obtained in 2016 at the ESO VLT-UT3, with the adaptive-optics camera SPHERE-IRDIS. 

We detected the trailing arcs Fraternit\'e and Egalit\'e, as well as the satellite Galatea for which we derived more accurate mean motion values.
 From the accurate determination of the mean motions, we confirm the mismatch between the arcs' position and the location of the 42:43 CIR and the CER with Galatea; thus showing that no 42:43 corotation model can explain the persistance of Neptune's incomplete ring arcs. 
 
Moreover, our photometric analysis of the arcs confirms that the leading arcs Courage and Libert\'e have faded away, while the trailing arcs Fraternit\'e and \'Egalit\'e are persistent and stable since their discovery.   

Regular imaging data, in particular with Galatea far from the arcs system, to avoid light contamination, are needed to follow the global time evolution of the system, and to propose a more global model for the arcs confinement that takes into account the disappearance of the two leading arcs Libert\'e and Courage.

\section{Acknowledgment}
 This work has made use of the SPHERE Data Centre, jointly operated by OSUG/IPAG (Grenoble), PYTHEAS/LAM/CeSAM (Marseille), OCA/Lagrange (Nice), Observatoire de Paris/LESIA (Paris), and Observatoire de Lyon (OSUL/CRAL), and supported by a grant from Labex OSUG@2020 (Investissements d’avenir – ANR10 LABX56).

This work was funded by the European Research Council under the European Community's H2020 (2014-2021/ERC Grant Agreement No. 669416). 

We thank the reviewer for their thorough revision of our paper and  constructive comments that significantly improve the quality of this work.

\bibliographystyle{aa}
\bibliography{references}
\end{document}